\documentclass[12pt]{article}

\usepackage{amsbsy}
\usepackage{amsmath}
\usepackage{amssymb}
\usepackage{amsthm}
\usepackage{authblk}
\usepackage{bbm}
\usepackage{color}
\usepackage{enumitem}
\usepackage{geometry}
\usepackage{graphicx}
\usepackage{natbib}
\usepackage[hidelinks]{hyperref}
\usepackage{mathtools}
\usepackage{setspace}
\usepackage{enumerate}
\usepackage{natbib}

\usepackage{bm}

\newtheorem{theorem}{Theorem}
\newtheorem{lemma}{Lemma}

\newtheorem{remark}{Remark}
\newtheorem{prop}{Proposition}
\newtheorem{definition}{Definition}
\newtheorem{Condition}{Condition}

\def\proof{{\it Proof.\ }}

\def\FF{\mathbb{F}}
\def\RR{\mathbb{R}}

\def\mathB{\mathcal{B}}

\def\mathX{\mathcal{X}}

\def\epf{\quad $\blacksquare$}

\begin{document}
\title{Shape-restricted transfer learning analysis for generalized linear regression models}
\author{Pengfei Li$^1$, Tao Yu$^2$, Chixiang Chen$^3$,  and Jing Qin$^4$}

\date{}
\maketitle
\vspace{-0.36in}

\begin{center}
$^1$Pengfei Li is Professor,
Department of Statistics and Actuarial Sciences,
University of Waterloo,
Waterloo, ON, Canada, N2L 3G1\\
(Email: \emph{pengfei.li@uwaterloo.ca})\\
$^2$Tao Yu is Associate Professor,
Department of Statistics and Data Science,
National University of Singapore, Singapore, 117546\\
(Email: \emph{yu.tao@nus.edu.sg})
\\
$^3$Chixiang Chen is 
Assistant professor, Department of Epidemiology and Public Health,
University of Maryland School of Medicine, MD 21201, U.S.A.\\
 (Email: \emph{Chixiang.Chen@som.umaryland.edu})\\
$^4$Jing Qin is Mathematical Statistician, National Institute of Allergy and Infectious Diseases, National Institutes of Health, MD 20892, U.S.A.\\
 (Email: \emph{jingqin@niaid.nih.gov})\\
\end{center}

\begin{abstract}

Transfer learning has emerged as a highly sought-after and actively pursued research area within the statistical community. The core concept of transfer learning involves leveraging insights and information from auxiliary datasets to enhance the analysis of the primary dataset of interest.
In this paper, our focus is on datasets originating from distinct yet interconnected distributions. We assume that the training data conforms to a standard generalized linear model, while the testing data exhibit a connection to the training data based on a prior probability shift assumption. Ultimately, we discover that the two-sample conditional means are interrelated through an unknown, nondecreasing function. We integrate the power of generalized estimating equations with the shape-restricted score function, creating a robust framework for improved inference regarding the underlying parameters.
We theoretically establish the asymptotic properties of our estimator and demonstrate, through simulation studies, that our method yields more accurate parameter estimates compared to those based solely on the testing or training data. Finally, we apply our method to a real-world example.


\end{abstract}

\noindent%
{\it Keywords:} Estimating equation; Generalized linear model; Prior Probability shift; Shape-restricted inference; Transfer learning

\section{Introduction} \label{Section-Intro}

In the era of big data and diverse sources of information, statistical transfer learning has emerged as a powerful paradigm to address challenges posed by data heterogeneity and limited sample sizes. Traditional statistical methods often assume that the training and testing data are drawn from the same distribution \citep{Imbens1994, Qin2000}, an assumption that may not hold true in many real-world scenarios. 
The motivation behind statistical transfer learning stems from the recognition that data collected from one domain can contain valuable information benefiting the learning process in another domain \citep{Weiss2016, Zhuang2020}. This is particularly relevant when the target domain has limited labeled data or exhibits different statistical properties compared to the source domain. By harnessing the wealth of knowledge from related domains, statistical transfer learning offers a way to enhance predictive accuracy, generalization, and interpretability in various applications, such as the development of precision medicine  \citep{Hodson2016}, language recognition \citep{Huang2013}, medical diagnosis \citep{Hajiramezanali2018}, and recommender systems \citep{Pan2013}.

The predictive accuracy of machine learning models is significantly impacted by the presence of data shift, which occurs when the distribution of the data used for model training differs substantially from the distribution of the data encountered during deployment. This issue is particularly pronounced in clinical AI systems, where machine learning algorithms play a crucial role in discerning patterns from clinical data. Dataset shift can lead to severe consequences, adversely affecting model performance and patient outcomes in such systems.
The impact of dataset shift becomes notably apparent in the case of the sepsis-alerting model at the University of Michigan Hospital \citep{Finlayson2021}. 
Developed by Epic Systems, this model effectively identified potential sepsis cases using historical patient data.
However, during the coronavirus disease 2019 pandemic, the model's performance deteriorated due to shifts in patients' demographic characteristics and other relevant factors.
This dataset shift fundamentally altered the relationship between fevers and bacterial sepsis, resulting in spurious alerting. As a consequence, the hospital's clinical AI governing committee deactivated the model for patient safety reasons.
Therefore, it is crucial to explore methods that can appropriately combine data information from similar studies when conducting statistical learning across various scientific fields, including economics, machine learning, distributed statistical inferences, and others.


Data shift can manifest in various forms. One extensively studied transfer learning problem involves binary classification within the framework of domain adaptation. In this scenario, the distributions of data pairs 
$(X, Y)$ are assumed to be drawn from related but distinct distributions defined on $\RR^p\times\{0,1\}$. 
For simplicity, we focus on 
two distributions--namely, the training and testing distributions. In this framework, two general strategies are employed. The first strategy involves making no additional assumptions about the relationship between distributions but instead suggesting a measurement of the divergence of the distributions. This measurement is then used to establish and quantify the effectiveness of the classification rules  \citep{Bendavid2007, Mansour2009}. The second strategy involves imposing structural assumptions between the training and testing distributions. Common assumptions include covariate shift and label shift (or more generally, prior probability shift). 
The covariate shift refers to the change in the distribution of the covariates $X$ in the training and testing data \citep{Shimodaira2000, Kpotufe2018, caiLiLiu2022}. Label shift assumes that the distributions of the covariates $X$ may remain similar, while the distributions of $Y|X$ can be significantly different \citep{caiwei2021, Maity2021}. Under this assumption, \citet{Obst2021} considered the linear regression model, i.e., $Y = X^T\beta + \epsilon$, when $Y$ is a continuous variable. In their approach, they proposed to estimate $\beta$ by a weighted sum of the corresponding least square estimates of $\beta$ from the training and testing data. By assuming that the regression coefficients are different, but model structures are the same, for training and testing data \citet{Li2022} and \citet{Tian2021} considered the transfer learning method for high dimensional linear models and generalized linear models with a Lasso penalty. 

The transfer learning method for the generalized linear model in the context of ``prior probability shift" has not been thoroughly explored in the literature. This paper aims to introduce a method under this setup that can effectively address the heterogeneity between the training and testing data. 
Specifically, we assume that the distributional difference between the training data and the testing data lies in the conditional distributions of $Y|X$, which is captured by the conditional density relationship:
\begin{equation}
f^*(y|\bm{x}) \propto \pi(y) f(y|\bm{x}), \label{eq-intro-1}
\end{equation}
where $\pi(y)$ is an unknown selection function, and $f(y|\bm{x})$ and $f^*(y|\bm{x})$ are the conditional densities of $Y|X = \bm{x}$ for the training and testing data, respectively.
In general, there is no strict requirement for the marginal densities of $X$ in the training and testing data to be identical. Consequently, the conditional densities $f(\bm{x} | y)$ in the training and testing data might show variations. Essentially, our assumption is somewhat less stringent than the previously mentioned probability shift assumption.

We observe that Assumption \eqref{eq-intro-1} is well-motivated in practical scenarios. Let $D=1$ or $0$ denote whether an individual data point is observed or not. If
\[
P(D=1|X = \bm{x},Y=y)=P(D=1|Y=y)=\pi(y)
\]
depending on the outcome $Y$ only, then, by Bayes's formula, we can show that
\begin{equation}
\label{model.selection}
Y|\{X = \bm{x}, D=1\}
\sim \frac{\pi(y)f(y|\bm{x})}
{\int \pi(y)f(y|\bm{x})dy}.
\end{equation}
In other words, if the selection bias function depends solely on the outcome, it leads to a scenario resembling a prior probability shift issue. 
It's worth noting that model \eqref{eq-intro-1} demonstrates a high degree of flexibility. It accommodates situations where the testing data stem from a selection-biased variant of the training data. Furthermore, it allows for differences in covariate densities between the training and testing data. In practical applications, such as clinical trials conducted across various research institutes, variations in patient inclusion criteria can result in distinct covariate distributions.


The issue known as the prior probability shift problem transforms into what is commonly referred to as the label shift problem when the outcome $Y$ is discrete. For example, when $Y$ represents an individual's reported disease status and $X$ encompasses symptoms and medical test outcomes obtained from standard equipment, this challenge is characterized as the label shift problem.
In some instances, it may be reasonable to assume that the distribution of $X$ is consistent across various studies. However, given a specific set of observations $X = \bm{x}$, the reported disease status from these studies may be influenced by $Y$. This influence can be attributed to various factors, such as differences in the knowledge and education backgrounds of doctors or environmental impacts. 
The aforementioned sepsis-alerting model \citep{Finlayson2021} developed by the University of Michigan Hospital stands as a concrete example of this label shift phenomenon.

We assume that Equation \eqref{eq-intro-1} holds, and that observations in the training data follow the exponential family with 
$E(Y|X = \bm{x})=\mu(\bm{x}^T\beta)$. 
Under these assumptions, 
 we demonstrate  (refer to  Section \ref{Section-Method}) that for the testing data,
\begin{equation}
E(Y|X = \bm{x}) = \psi\left(\mu(\bm{x}^T \beta)\right), \label{eq-intro-2}
\end{equation}
with $\psi(\cdot)$ being a monotone function. This is inline with the development in \citet{Maity2021}, in which they assumed that the regression models for binary outcomes $Y$ in both the training and testing data are linked through a known nondecreasing link function and a low-dimensional transformation of the covariates. This innovative perspective draws inspiration from the Cox proportional hazard model.
In our current paper, we deviate from the assumption of a binary outcome and, instead, focus on more general regression functions in both the training and testing data. These functions are interconnected through an unknown monotone transformation, presenting an intriguing challenge. Essentially, this shift transforms our problem into a shape-restricted transfer learning scenario, paving the way for exciting avenues of exploration and investigation. Notably, shape-restricted inference has found applications in various domains, including AI \citep{Gupta2016, Wang2020}. In transfer learning scenarios, the incorporation of shape restrictions, such as monotonicity, proves particularly advantageous, as it enables us to align with prior knowledge and enforce domain-specific constraints.

It's worth noting that assuming a generalized linear model for the training data, Equation \eqref{model.selection} implies Equation \eqref{eq-intro-2}. However, Equation \eqref{eq-intro-2} presents a more general condition than Equation \eqref{model.selection}.
In particular, when the outcome variable $Y$ is binary, \citet{caiwei2021} and \citet{reeve2021adaptive} have made similar assumptions concerning a monotone link function for the transfer learning problem. Their primary focus has been on achieving minimax optimal rates of convergence in their respective studies.
This paper focuses on the assumption stated in Equation \eqref{eq-intro-2}.

The rest of the paper is organized as follows.  In Section \ref{Section-Method}, we present the model assumptions for both the training and testing data, and introduce our estimating equation-based method for $\beta$. This method incorporates information from both the training and testing data, assuming that $\psi(\cdot)$ in \eqref{eq-intro-2} is unknown.
In Section \ref{section-asymp}, we theoretically establish the asymptotic normality of our $\beta$ estimates. Section \ref{Section-Sim} presents simulation studies, while Section \ref{Section-RealData} applies our method to a real data example. Finally, Section \ref{Section-Dis} concludes the paper with a discussion. For ease of presentation, technical details are provided in the Appendix and supplementary material.

\section{Estimation Method} \label{Section-Method}

Consider a set of training data: $(Y_{1,i}, X_{1,i}), i=1, \ldots, n$, and a set of testing data: $(Y_{2,j}, X_{2,j})$, $j=1, \ldots, m$.
Assume that the training data consists of independent and identically distributed (i.i.d.) copies of $(Y_1, X_1)$, which follows a standard parametric generalized linear model \citep{mccullagh1989generalized}. Specifically, assume that the conditional density of $Y_1$ given the covariate $X_1 = \bm{x}$ belongs to a canonical exponential family:
\begin{eqnarray}
f(y|\bm{x}) = \exp\left\{ \frac{y \theta(\bm{x}) - b(\theta(\bm{x}))}{a(\phi) } +  c(y, \phi)\right\}. \label{eq-method-1}
\end{eqnarray}
Here, $a(\cdot), b(\cdot)$, and $c(\cdot, \cdot)$
are known functions,  $\theta(\cdot)$ is the canonical parameter, and $\phi$ is the dispersion parameter. 
In the context of the parametric generalized linear model, we have $E(Y_1|X_1=\bm{x}) = \mu_1(\bm{x}^T \beta)$, where $\mu_1(\cdot)$ is a known increasing function. 
Subsequently, the following relationships hold:
\begin{eqnarray*}
\mu_1(\bm{x}^T \beta) &=& E(Y_1|X_1=\bm{x}) = b'(\theta(\bm{x})), \nonumber \\
\text{var}(Y_1|X_1=\bm{x}) &=& a(\phi) b''(\theta(\bm{x})) \equiv a(\phi) \cdot V(\bm{x}^T \beta), \label{eq-method-2}
\end{eqnarray*}
where $V(\cdot)$ is the variance function with a known function form. 
It is straightforward to check 
\begin{eqnarray*}
\frac{\partial \log f(y|\bm{x})}{\partial \beta} = \frac{\mu_1'(\bm{x}^T \beta) }{a(\phi) V(\bm{x}^T \beta)} \bm{x}\left\{ y - \mu_1(\bm{x}^T \beta)\right\}.   \label{eq-method-3}
\end{eqnarray*}
Assuming that the dispersion parameter $\phi$ is a constant across the observations, the maximum likelihood estimate for $\beta$ can be obtain by solving the equations $\psi_{1,n}(\beta) = 0$, where 
\begin{eqnarray}
\psi_{1,n}(\beta) &=& \frac{1}{n} \sum_{i=1}^n \frac{\mu_1'(X_{1,i}^T\beta)}{V(X_{1,i}^T\beta)} X_{1,i} \left\{ Y_{1,i} - \mu_1(X_{1,i}^T\beta) \right\}.  \label{eq-method-4}
\end{eqnarray}

For the testing data $(Y_{2,j}, X_{2,j}), j=1,\ldots,m$, we assume that they are i.i.d. copies of $(Y_2, X_2)$. 
The conditional distribution of $Y_2$ is assumed to be related to that of the training data. Specifically, we assume that the conditional density of $Y_2|X_2 = \bm{x}$, denoted by $f^*(y|\bm{x})$, satisfies \eqref{eq-intro-1}, with $\pi(y)$ representing an unknown selection function. 
Consequently,
\begin{eqnarray}
f^*(y|\bm{x}) = \frac{\pi(y) f(y|\bm{x})}{\int \pi(y) f(y|\bm{x}) dy}. \label{eq-method-5}
\end{eqnarray}
The following proposition holds. 
\begin{prop} \label{proposition-1}
Assume the conditional density of $Y_2|X_2 = \bm{x}$ in \eqref{eq-method-5}, with $f(y|\bm{x})$ given by \eqref{eq-method-1}. Then, there exists a nondecreasing function $\mu_2(\cdot)$ such that
\begin{eqnarray}
E(Y_2|X_2 = \bm{x}) = \mu_2(\bm{x}^T \beta). \label{eq-prop-1-0}  
\end{eqnarray}
\end{prop}
\proof Combining \eqref{eq-method-1} and \eqref{eq-method-5}, we obtain 
\begin{eqnarray*}
 f^*(y|\bm{x})
 &=& \frac{\exp\left[ y \theta(\bm{x})/a(\phi)+\left\{\log \pi(y)+c(y,\phi)\right\} \right]}
 {\int \exp\left[ y \theta(\bm{x})/a(\phi)+\left\{\log \pi(y)+c(y,\phi)\right\} \right] dy}\\
 &=& \exp\left[ \frac{y \theta(\bm{x})-  b_\pi(\theta(\bm{x}), \phi)}{a(\phi)}+\left\{\log \pi(y)+c(y,\phi)\right\} \right],
\end{eqnarray*}
where 
\begin{eqnarray*}
b_\pi(\theta(\bm{x}), \phi) = a(\phi) \log \int \exp\left[ y \theta(\bm{x})/a(\phi)+\left\{\log \pi(y)+c(y,\phi)\right\} \right] dy. 
\end{eqnarray*}
Denote by $E_{Y_2|X_2 = \bm{x}}(\cdot)$ the expectation with respect to $Y_2$ under the density $f^*(y|\bm{x})$. 
Due to the fact $
E_{Y_2|X_2 = \bm{x}}\left[ {\partial \log f^*(Y_2|\bm{x})}\big/{\partial \left\{\theta(\bm{x}) \right\}} \right] = 0,$
we have
\begin{eqnarray*}
E_{Y_2|X_2 = \bm{x}}\left\{Y_2 - \frac{\partial b_\pi(\theta(\bm{x}), \phi)}{\partial \theta(\bm{x})}\right\} = 0,
\end{eqnarray*}
or equivalently,
\begin{eqnarray}
E(Y_2|X_2 = \bm{x}) = \frac{\partial b_\pi(\theta(\bm{x}), \phi)}{\partial \theta(\bm{x})}. \label{eq-prop-1-1}
\end{eqnarray}
Furthermore, based on the fact that 
\begin{eqnarray*}
E_{Y_2|X_2 = \bm{x}}\left[ \frac{\partial^2 \log f^*(Y_2|\bm{x})}{\partial \left\{\theta(\bm{x}) \right\}^2} \right] + E_{Y_2|X_2 = \bm{x}} \left(\left[ \frac{\partial \log f^*(Y_2|\bm{x})}{\partial \left\{\theta(\bm{x}) \right\}} \right]^2 \right) = 0,
\end{eqnarray*}
we have
\begin{eqnarray}
\frac{\partial^2 b_\pi(\theta(\bm{x}), \phi)}{\partial \left\{\theta(\bm{x})\right\}^2} &=& \frac{1}{a(\phi)} \cdot E_{Y_2|X_2 = \bm{x}}\left[\left\{Y_2 - \frac{\partial b_\pi(\theta(\bm{x}), \phi)}{\partial \theta(\bm{x})}\right\}^2\right] \nonumber \\
&=& \frac{\text{var}(Y_2|X_2 = \bm{x})}{a(\phi)} \geq 0, \label{eq-prop-1-2}
\end{eqnarray}
where ``$=$" holds if and only if $Y_2 = E(Y_2|X_2 = \bm{x})$ almost surely. Combining \eqref{eq-prop-1-1} and \eqref{eq-prop-1-2}, we conclude that $E(Y_2|X_2 = \bm{x})$ is a nondecreasing function of $\theta(\bm{x})$, and therefore is a nondecreasing function of $\bm{x}^T \beta$.  We complete the proof of the proposition. \epf

With Proposition \ref{proposition-1}, the testing data can be analyzed using the monotone single index model \citep{Balabdaoui2017, Balabdaoui2019}. Specifically, we can employ the following profiling procedure to establish a set of estimating functions based on the testing data. For each given $\beta$, we estimate $\mu_2(\cdot)$ by
\begin{eqnarray}
\widehat \mu_{2,\beta}(\cdot) = \arg\min_{\mu_2} \sum_{j=1}^m \left\{Y_{2,j} - \mu_2\left(X_{2,j}^T\beta\right)\right\}^2, \label{eq-method-6}
\end{eqnarray}
subject to $\mu_2(\cdot)$ being a nondecreasing function. 
This can be solved by the standard PAVA algorithm \citep{Ayer1955}. Employing similar ideas as \citet{Groeneboom2018} and \citet{Yu2023}, we can construct a set of estimating functions:
\begin{eqnarray}
\psi_{2,m}(\beta) = \frac{1}{m}\sum_{j=1}^m X_{2,j}\left\{Y_{2,j}-\widehat \mu_{2,\beta}\left(X_{2,j}^T\beta\right)\right\}.  \label{eq-method-7}
\end{eqnarray}
Based only on the testing data, a $\beta$ estimate can be obtained by solving the equation $\psi_{2,m}(\beta) = 0$. 

Combining \eqref{eq-method-4} with \eqref{eq-method-7}, let 
\begin{eqnarray}
\psi_{n,m}(\beta) = \lambda_{n,m} \psi_{1,n}(\beta) + (1-\lambda_{n,m}) \psi_{2,m}(\beta), \label{eq-method-8}
\end{eqnarray}
where $\lambda_{n,m} = n/(n+m)$. We hope to define $\widehat \beta$ to be the root of $\psi_{n,m}(\beta) = 0$. However, $\psi_{n,m}(\cdot)$ may not be a continuous mapping, and its root may not exist. Similar to \citet{Groeneboom2018}, we define $\widehat \beta$ to be the {\it zero crossing} of $\psi_{n,m}(\beta)$. The zero crossing of a function (or a mapping) is defined below.

\begin{definition}
For a function $\psi$: $\mathX \to \RR$, $\bm{x}$ is called the zero-crossing of $\psi$ if every open neighborhood of $\bm{x}$ contains $\bm{x}_1, \bm{x}_2$ such that $\psi(\bm{x}_1)\psi(\bm{x}_2) \leq 0$.
For a mapping $\widetilde \psi$: $\mathX \to \RR^d$ for $d\geq 2$, $\bm{x}$ is called the zero-crossing of $\widetilde \psi$ if $\bm{x}$ is the zero-crossing of each component of $\widetilde \psi$.
\end{definition}

In most practical scenarios, it is reasonable to assume that the size of the training data, i.e., $n$, is less than or comparable to that of the testing data, i.e., $m$; in the sense $\lambda_{n,m} \to \lambda > 0$. Some discussion on this assumption can be found in \citet{caiwei2021} and \citet{Maity2021}. However, from both a methodological and theoretical standpoint, it is possible for $n \ll m$ or, more rigorously, for $n/m \to 0$, resulting in $\lambda_{n,m} \to \lambda = 0$. Intuitively, when $\lambda_{n,m}$ is small, in other words, when we have much more testing data than training data, the contribution of the training data to the $\beta$ estimation (through $\psi_{1,n}(\beta)$) is limited; therefore, $\widehat{\beta}$ relies mainly on the testing data (through $\psi_{2,m}(\beta)$).
On the other hand, with only the testing data, model \eqref{eq-prop-1-0} is identifiable only when a constraint is imposed on $\beta$, such as
$\|\beta\|_2 = 1$. Theoretically, the zero crossing of $\psi_{2,m}(\beta)$ only exists asymptotically for $p-1$ out of its $p$ components, such as the 2nd to $p$th component if $\beta_{0,1} \neq 0$, where $\beta_0$ denotes the true value of $\beta$, and $\beta_{0,1}$ denotes the first component of $\beta_0$. See Section \ref{section-asymp} and the corresponding technical developments in the supplementary material for details.
Practically, when $\lambda_{n,m}$ is very small, we suggest estimating $\widehat{\beta}$ as the zero crossing of  $\psi_{n,m}(\beta)_{2:p}$ under the constraint $\|\beta\|_2 = 1$, where $\psi_{n,m}(\beta)_{2:p}$ is a vector composed of the 2nd to $p$th component of $\psi_{n,m}(\beta)$.

\section{Asymptotic Properties} \label{section-asymp}

In this section, we will establish the asymptotic properties of $\widehat \beta$. Besides the notations in Section \ref{Section-Method}, we need some additional notations. Let $\beta_0$ and $\mu_{2,0}(\cdot)$ denote the true values of $\beta$ and $\mu_2(\cdot)$, respectively. Also, let $\varphi_\beta(t) = E\left(X_2|X_2^T\beta = t\right)$, $\mu_{2, \beta}(t) = E\left(Y_2|X_2^T\beta = t\right)$; thus $\mu_{2,0}(\cdot) = \mu_{2,\beta_0}(t)$.  
Denote
\begin{eqnarray*}
A(\beta_0)  
&=&  \lambda E\left[ \frac{\left\{\mu_1'(X_1^T\beta_0)\right\}^2}{V(X_1^T \beta_0)} X_1 X_1^T \right] \\
&& + (1-\lambda)E\left[ \left\{X_2 - \varphi_{\beta_0}(X_2^T \beta_0)\right\}\left\{X_2 - \varphi_{\beta_0}(X_2^T \beta_0)\right\}^T \mu_{2,0}'\left(X_2^T\beta_0 \right)   \right], \\
\Sigma_1 &=& \text{var}\left[ \frac{\mu_1'(X_1^T\beta_0)}{V(X_1^T\beta_0)} X_1 \left\{ Y_1 - \mu_1(X_1^T \beta_0) \right\} \right],\\
\Sigma_2 &=& \text{var}\left[ \left\{X_2 - \varphi_{\beta_0}\left(X_2^T \beta_0 \right) \right\} \left\{Y_2 - \mu_{2,0}\left(X_2^T\beta_0\right)\right\} \right],
\end{eqnarray*}
where $\mu'(t) = {\partial \mu(t)}/{\partial t}$ for $\mu(t)$ being a function differentiable at $t$. Furthermore, let 
\begin{eqnarray*}
\lim_{n\to \infty, m \to \infty} \lambda_{n,m} = \lambda.
\end{eqnarray*}
As discussed earlier, we need to consider two possibilities: $\lambda > 0$ and $\lambda = 0$. Let us first consider the case where $\lambda >0$. Note that this includes $\lambda = 1$ as a special case, which implies that we have much more training data than testing data.

We impose the following regularity conditions in the development of the theoretical results. They are not necessarily the weakest possible.

\begin{Condition} \label{Condition-1}
$\beta \in \mathB$, $F_{X_1}(\bm{x})$ and $F_{X_2}(\bm{x})$ are supported on $\mathX$, where both $\mathB$ and $\mathX$ are compact subspaces of $\RR^p$, and $F_{X_s}(\cdot)$ for $s=1$ or 2 denotes the cumulative distribution function (c.d.f.) of $X_s$.  
\end{Condition}

\begin{Condition} \label{Condition-2}
There exists a universal constant $C>0$, such that for any $\beta_1, \beta_2 \in \mathB$, 
\begin{eqnarray*}
\sup_{t\in \RR} \int \left| I\left(\beta_1^T \bm{x} \leq t\right) - I\left(\beta_2^T \bm{x} \leq t \right)  \right| d F_{X_2}(\bm{x}) \leq C \|\beta_1 - \beta_2\|_1,
\end{eqnarray*}
where $\|\cdot \|_q$ denotes the $l_q$ norm. 
\end{Condition}

\begin{Condition} \label{Condition-3}
The conditional probability density function $f_{X_2|X_2^T\beta = t }(\bm{x})$ is continuously differentiable in $\beta$ and $t$. 
\end{Condition}

\begin{Condition} \label{Condition-5}

$\mu_{2,\beta}(t)$ is continuously differentiable for both $\beta\in \mathB$ and $t\in  \{\bm{x}^T\beta: \bm{x}\in \mathX, \beta \in \mathB\}$, and there exists a universal constant $\eta_0 >0$, such that 
\begin{eqnarray*}
\inf_{t\in \left\{\bm{x}^T\beta: \bm{x}\in \mathX, \beta \in \mathB \right\}, \|\beta - \beta_0\|_2 \leq \eta_0} \frac{\partial \mu_{2,\beta}(t)}{\partial t}> 0. 
\end{eqnarray*}
\end{Condition}

\begin{Condition} \label{Condition-6}
The density function of $X_1$, denoted by $f_{X_1}(\bm{x})$, is continuously differentiable for $\bm{x}\in \mathX$, and $E(X_1X_1^T)$ is of full rank.  
\end{Condition}

\begin{Condition} \label{Condition-4}
For $s=1$ or 2, and any integer $k\geq 2$, $E\left(|Y_s|^k|X_s\right) \leq k! M^{k-2} v/2$ for some universal constants $M$ and $v$.
\end{Condition}

\begin{Condition} \label{Condition-7}
For $t \in \left\{\bm{x}^T\beta: \bm{x} \in \mathX, \beta \in \mathB \right\}$, $\mu_1(t)$ is nondecreasing and continuously differentiable, $\mu_1'(t)$ is bounded away from zero and infinity; and $V(t)$ is continuously differentiable and is bounded away from 0 and $\infty$.  
\end{Condition}

\begin{Condition} \label{Condition-8}
For any $\beta_1, \beta_2\in \mathB$, such that $\beta_1 \neq c \beta_2$, for all $c \in \RR$, and $c\neq 0$, we have, 
\begin{eqnarray*}
E\left\{\text{var}\left( \beta_1^T X_2 | \beta_2^T X_2 \right)\right\} >0.
\end{eqnarray*}
\end{Condition} 

\begin{remark} \label{remark-identifiability}
Note that Condition \ref{Condition-8} implies that if $\beta_1 \neq c \beta_2$, for all $c \in \RR$, and $c\neq 0$, then,
\begin{eqnarray*}
P\left(\beta_1^T X_2 \neq  E\left( \beta_1^T X_2 | \beta_2^T X_2 \right) \right) >0;
\end{eqnarray*}
since otherwise, $\beta_1^T X_2 = E\left( \beta_1^T X_2 | \beta_2^T X_2 \right)$ almost surely, which implies 
\begin{eqnarray*}
\text{var}\left( \beta_1^T X_2 | \beta_2^T X_2 \right) =  E\left[\left\{\beta_1^T X_2 -  E\left( \beta_1^T X_2 | \beta_2^T X_2 \right)\right\}^2 \Big| \beta_2^T X_2 \right] = 0,
\end{eqnarray*}
almost surely, which contradicts Condition \ref{Condition-8}. 

\end{remark}

Theorem \ref{theorem-lambda-positive} below establishes the existency and asymptotic properties of $\widehat \beta$ when $\lambda >0$. 

\begin{theorem} \label{theorem-lambda-positive}
Assume Conditions \ref{Condition-1}--\ref{Condition-8}. 
Denote by $\widehat \beta$ the zero crossing of $\psi_{n,m}(\beta)$ (if it exists).
We have the following:
\begin{itemize}
\item[(1)] When $n\to \infty$ and $m \to \infty$, a zero crossing $\widehat \beta$ of $\psi_{n,m}(\beta)$ exists with probability tending to 1. 
\item[(2)] If $\widehat \beta$ is a zero crossing of $\psi_{n,m}(\beta)$, we have $\widehat \beta \to \beta_0$ in probability.
\item[(3)] For $\widehat \beta$ being a zero crossing of $\psi_{n,m}(\beta)$, we have
\begin{eqnarray*}
\sqrt{n+m} \left( \widehat \beta - \beta_0 \right) \to  A_0(\beta_0)^{-1} \cdot N\left(0, \lambda \Sigma_1 + (1-\lambda) \Sigma_2   \right), 
\end{eqnarray*}
in distribution. 
\end{itemize}
\end{theorem}

Next, we consider that $\lambda = 0$. In this case, we shall define $\widehat \beta$ to be the zero crossing of $\psi_{n,m}(\beta)_{2:p}$ under the constraint $\|\beta\|_2 = 1$; here for a vector $\bm{x}$, $\bm{x}_{i:j}$ denotes the vector formed of the $i$th to $j$th entries of $\bm{x}$. Theorem \ref{theorem-lambda-0} below establishes the asymptotic properties of this $\widehat \beta$. We need the following notation.
\begin{eqnarray*}
\widetilde A(\beta_0)  
&=& E\left[ \left\{X_2 - \varphi_{\beta_0}(X_2^T \beta_0)\right\}_{2:p}\left\{X_2 - \varphi_{\beta_0}(X_2^T \beta_0)\right\}_{2:p}^T \mu_{2, 0}'\left(X_2^T\beta_0 \right)   \right] \nonumber \\ 
&&\times \left(\beta_{0,2:p} \beta_{0,2:p}^T + I_{p-1}\right),\\
\widetilde \Sigma_2 &=& \text{var}\left[ \left\{(X_2)_{2:p} - \varphi_{\beta_0}\left(X_2^T \beta_0 \right)_{2:p} \right\} \left\{Y_2 - \mu_{2,0}\left(X_2^T\beta_0\right)\right\} \right]. 
\end{eqnarray*}

\begin{theorem} \label{theorem-lambda-0}
Assume Conditions \ref{Condition-1}--\ref{Condition-8}. Suppose $\lambda = 0$ and $\beta_{0,1} \neq 0$. 
Denote by $\widehat \beta$ the zero crossing of $\psi_{n,m}(\beta)_{2:p}$ (if it exists) under the constraint $\|\beta\|_2 = 1$.
We have the following:
\begin{itemize}
\item[(1)]  There exists a universal constant $c>0$, such that when $n\to \infty$ and $m \to \infty$, a zero crossing $\widehat \beta$ of $\psi_{n,m}(\beta)_{2:p}$ with $\widehat \beta_1 > c$ exists with probability tending to 1. 
\item[(2)] Assume that $\widehat \beta$ is a zero crossing of $\psi_{n,m}(\beta)_{2:p}$ such that $\widehat \beta_1 > c$ for a universal constant $c>0$, we have $\widehat \beta \to \beta_0$ in probability.
\item[(3)] For $\widehat \beta$ given in Part (2), we have
\begin{eqnarray*}
\sqrt{n+m} \left(\widehat \beta_{2:p} - \beta_{0, 2:p} \right)\to \widetilde A(\beta_0)^{-1} N(0, \widetilde \Sigma_2),
\end{eqnarray*}
in distribution.
\end{itemize}

\end{theorem}

For presentational continuity, we sketch the proof for Part (3) of Theorems \ref{theorem-lambda-positive} and \ref{theorem-lambda-0} in the Appendix, and relegate the lengthy technical details to the supplementary material.

\section{Simulation Study} \label{Section-Sim}
\subsection{Data Simulation}
We consider the following simulation setups. For both the training and testing data, we generate the covariate $X_{1,i}$ and $X_{2,j}$ for all $i$ and $j$ to be i.i.d. copies of $X = (U_1, U_2, U_3)^T$, where
\begin{eqnarray*}
U_1\sim \text{Unif}[-1.5,1.5],\quad U_2\sim \text{Bernoulli}(0.5), \quad U_3\sim N(0,1), 
\end{eqnarray*}
where $U_1$, $U_2$, and $U_3$ are independent. Given $X = \bm{x} \equiv (u_1, u_2, u_3)^T$, the responses of the training and testing data are then simulated based on two models. 
\begin{itemize}
\item {\bf Model 1 (Gamma model)}

The training data are simulated from $f(y|\bm{x})$, which is the Gamma distribution with mean $\exp(u_1 + u_2 + u_3)$ and shape parameter equal to 2. The testing data are simulated from $f^*(y|\bm{x}) \propto \pi(y) f(y|\bm{x})$ with 
\begin{eqnarray*}
\pi(y)=\frac{\exp(-0.5y)}{1+\exp(-0.5y)}.
\end{eqnarray*}

\item{\bf Model 2 (Normal model)}

The training data are simulated from
\begin{eqnarray*}
Y=2+u_1+u_2+u_3+\epsilon, ~~ \epsilon\sim N(0,1). 
\end{eqnarray*}
Therefore, $f(y|\bm{x})$ is the density function of $N(2+u_1 + u_2+u_3, 1)$. 
The testing data are simulated from $f^*(y|\bm{x}) \propto \pi(y) f(y|\bm{x})$ with 
\begin{eqnarray*}
\pi(y)=\frac{\exp(-0.5y)}{1+\exp(-0.5y)}.
\end{eqnarray*}
\end{itemize}

Thus, for both models, we have $\beta_0 = (1,1,1)^T$. We examine three combinations of sample sizes:  $(750,250)$, $(500,500)$, and $(250,750)$. For each model configuration, we perform 1000 repetitions. 

\subsection{Results}

To illustrate the advantages of our method, which combines training and testing data to improve $\beta$ estimation, we compare it to two other approaches: using only training data and using only testing data. Specifically, we consider 

\begin{itemize}
    \item the maximum likelihood estimate based only on the training data, called MLE estimate;
    \item the estimating equation estimate based only on the testing data, called EE estimate;
    \item the estimate based on the estimating equations \eqref{eq-method-8}, which combine the training data and the testing data, called our estimate. 
\end{itemize}

To obtain the EE estimate, we impose the constraint $\|\beta\|_2 = 1$. Therefore, to compare the performance of the methods, we standardise the estimates by
\begin{eqnarray*}
\eta_i=\frac{\beta_i}{\sqrt{\beta_1^2+\beta_2^2+\beta_3^2}}, \qquad i=1,2,3.
\end{eqnarray*}
The true value of  $(\eta_1,\eta_2,\eta_3)$ is given by  $\left(1/\sqrt{3},1/\sqrt{3},1/\sqrt{3}\right)$. 

We use the relative bias (RB) and mean square error (MSE) as the comparison criteria. For a true parameter $\eta_0$ and the corresponding estimates $\widehat \eta_1, \ldots \widehat \eta_N$ based on $N$ repetitions, its RB in percentage and MSE are defined to be
\begin{eqnarray*}
\text{RB} = \frac{\sum_{i=1}^N \left(\widehat \eta_i - \eta_0 \right)}{N}\times 100, \quad \text{MSE} = \frac{\sum_{i=1}^N \left(\widehat \eta_i - \eta_0 \right)^2 }{N}. 
\end{eqnarray*}

Table \ref{Table-1} presents the RBs and MSEs of the parameter estimates obtained over 1000 repetitions. Our method, which combines information from both the training and testing data, consistently produces the smallest MSEs for all model configurations. When the sample size of the training data is much larger than that of the testing data, specifically with $n=750$, $m=250$, and $\lambda_{n,m} = 0.75$, our method yields significantly smaller MSEs compared to the EE method and slightly smaller MSEs than the MLE method. When $n=250$ and $m=750$, and $\lambda_{n,m} = 0.25$, our method achieves much smaller MSEs than both the MLE and EE methods. These results underscore the importance of transfer learning methods in the analysis of testing data, when one or more sets of training data are available. Our method effectively incorporates information from the training data to improve estimation and inference for the testing data.

\begin{table}[ht]
{ \tabcolsep=1mm
\begin{center}
\caption{Relative bias (\%) and MSE ($\times 1000$) for  $(\eta_1,\eta_2,\eta_3)$}
\medskip
\begin{tabular}{cc|ccc|ccc|ccc}
\hline
$(n,m)$&&\multicolumn{3}{c|}{MLE}&\multicolumn{3}{c|}{EE}&
\multicolumn{3}{c}{ Our}\\
&& $\eta_1$ & $\eta_2$ & $\eta_3$& $\eta_1$ & $\eta_2$ & $\eta_3$& $\eta_1$ & $\eta_2$ & $\eta_3$\\
\hline
&&\multicolumn{9}{c}{Model 1: Gamma model}\\

(750,250)&RB &0.04&-0.15&-0.03&-0.85&1.07&-1.48&0.01&-0.05&-0.09\\
        &MSE &0.24&0.46&0.24&2.09&4.07&2.21&0.22&0.41&0.23\\
(500,500)&RB &0.06&-0.22&-0.04&-0.54&1.01&-1.11&-0.03&0.02&-0.15\\
        &MSE &0.36&0.66&0.35&1.11&2.08&1.09&0.29&0.50&0.29\\
(250, 750)&RB   &0.00&-0.36&-0.07&-0.48&0.95&-0.89&-0.13&0.17&-0.26\\
        &MSE&0.75&1.37&0.71&0.70&1.36&0.76&0.38&0.66&0.38\\
\hline
&&\multicolumn{9}{c}{Model 2: Normal model}\\
(750,250)&RB   &0.05&-0.25&-0.07&-1.30&2.17&-1.89&-0.05&-0.02&-0.15\\
        &MSE&0.49&0.89&0.45&1.91&3.28&1.63&0.39&0.72&0.36\\
(500,500)&RB   &0.07&-0.39&-0.09&-0.82&1.41&-1.12&-0.09&0.03&-0.17\\
        &MSE&0.74&1.29&0.66&0.96&1.70&0.82&0.41&0.72&0.37\\
(250,750)&RB   &-0.09&-0.30&-0.45&-0.51&0.99&-0.81&-0.13&0.17&-0.27\\
        &MSE&1.58&2.66&1.35&0.61&1.07&0.52&0.41&0.74&0.40\\
\hline
\end{tabular} \label{Table-1}
\end{center}
}
\end{table}

\section{A Real Data Example} \label{Section-RealData}

In this section, we apply our proposed method to study potential risk factors associated with blood pressure for the subcohort of white males whose ages are greater than or equal to 45 and less than 55 (in the range of middle age).

We consider two datasets: the UK Biobank data as the testing data and the data from the Atherosclerosis Risk in Communities (ARIC) study \citep{Wright2021} as the training data. The UK Biobank is a remarkable resource that has significantly contributed to the field of biomedical research. The database's extensive collection of genetic and health information from a diverse group of half a million UK participants provides researchers with a wealth of data to explore and analyze. Its ongoing expansion and accessibility to approved researchers have facilitated crucial investigations into various diseases and health conditions.
The ARIC study by the National Heart, Lung, and Blood Institute is a comprehensive research initiative with the primary objective of delving into the causes and clinical consequences of atherosclerosis. The ARIC study focuses on studying a diverse group of adults residing in four distinct communities within the United States. By collecting extensive data on these individuals, researchers aim to uncover the underlying factors that contribute to the development and progression of atherosclerosis. This includes investigating genetic, environmental, and lifestyle factors that may influence an individual's susceptibility to atherosclerosis and its associated complications.


Both datasets appear to include several key variables related to health and risk factors. These variables are crucial for studying various aspects of health and disease. Here's a brief summary of the mentioned variables:
\begin{itemize}
\item 
Systolic Blood Pressure (SBP): This variable measures the pressure in the arteries when the heart beats. It is typically measured in millimeters of mercury ($mmHg$) and is an important indicator of cardiovascular health.

\item Use of Anti-Hypertensive Medication (HYMD): This binary variable (1 for yes, 0 for no) indicates if the individual is taking medication to manage high blood pressure (hypertension).

\item Glucose Levels (Glucos): This variable measures glucose levels in the blood, often using NMR metabolomics. Glucose is a primary energy source for the body and plays a crucial role in diabetes and metabolic health.

\item Age (Age): Age is a fundamental demographic variable. It is often considered a significant factor in various health outcomes as many conditions are age-dependent.

\item Body Mass Index (BMI): BMI is a measure of body fat based on an individual's weight and height. It is commonly used to assess weight-related health risks.

\item Smoker (Smoker): This binary variable (1 for yes, 0 for no) indicates whether an individual smokes. Smoking is a known risk factor for various health issues, including cardiovascular disease and cancer.
\end{itemize}

The mention of data cleaning and filtering suggests that the datasets have been processed to remove errors or outliers, ensuring the quality and reliability of the data for research and analysis. Additionally, it's worth noting the sample sizes for each dataset: the UKBiobank dataset has 729 observations, while the ARIC dataset has 786 observations. These observations likely represent unique individuals, and researchers can use this data to explore relationships between these variables and various health outcomes.

As highlighted in the introduction section, model (\ref{eq-intro-1}) demonstrates remarkable flexibility. It accommodates the presence of selection bias and variations in covariate distributions, which are often attributed to differences in patient inclusion criteria between studies conducted in the UK and the USA. As a consequence, employing model (\ref{eq-intro-1}) would provide a more reliable approach than assuming that the UK Biobank data and ARIC data share an identical joint distribution.

We use $\log(\text{SBP})$ as the continuous response and five other variables as risk factors. It is highly plausible that a prior probability shift issue exists between the UKBiobank and ARIC data due to different study designs and studied populations (USA vs Europe): the averages of SBP in UKBiobank and ARIC data are 135.5 and 115.6, respectively. We apply the three methods from our simulation study to analyze the two datasets: the MLE estimate based on the linear model for the ARIC data (training data); the EE estimate based on the UKBiobank data (testing data); the proposed estimate using both datasets. Table \ref{UK.ARIC} summarizes the point estimate, the corresponding bootstrap standard error (SE), and the 95\% bootstrap percentile confidence interval (BPCI). For comparison, the coefficients for the five risk factors are standardized to have a length of 1.

\begin{table}[!ht]
\begin{center}
\caption{Analysis results for the ARIC and UKBiobank data\label{UK.ARIC}}
\begin{tabular}{lccc}
\hline
Variable&Point Estimate&Bootstrap SE&95\% BPCI\\
\hline
&\multicolumn{3}{c}{MLE estimate based on linear model for the ARIC data}\\
HYMD     &0.855&0.225&(0.135,0.978)\\
Glucos   &0.318&0.180&(0.072,0.754)\\
Age      &0.143&0.060&(0.048,0.280)\\
BMI      &0.159&0.076&(0.069,0.316)\\
Smoker   &-0.350&0.309&(-0.837,0.291)\\
\hline
&\multicolumn{3}{c}{EE method based on the UKBiobank data}\\
HYMD    &-0.197&0.587&(-0.696,0.999)\\
Glucos  &0.927&0.333&(0.024,0.964)\\
Age     &0.101&0.048&(0.000,0.157)\\
BMI     &0.270&0.090&(0.009,0.305)\\
Smoker  &-0.139&0.281&(-0.855,0.016)\\
\hline
&\multicolumn{3}{c}{Proposed method based on both data}\\
HYMD   &0.900&0.177&(0.398,0.975)\\
Glucos &0.332&0.151&(0.133,0.696)\\
Age    &0.138&0.049&(0.046,0.239)\\
BMI    &0.183&0.055&(0.079,0.283)\\
Smoker &-0.167&0.250&(-0.681,0.245)\\

\hline
\end{tabular}
\end{center}
\end{table}

In Table \ref{UK.ARIC}, we noticed that the proposed method leads to similar point estimates and detects the same significant risk factors at the 5\% level as the MLE method based on the linear model for the ARIC data only, but with much smaller bootstrap SE and shorter BPCIs. The risk factors include HYMD, Glucos, Age, and BMI. In contrast, the EE method based on the UKBiobank data only fails to identify the risk factor HYMD at the 5\% level, and furthermore, the lower bound of the 95\% CI for the coefficient of the Age variable is close to 0. These results together show the advantage of the proposed method.

\begin{figure}[!ht]
\centering{\includegraphics[scale=0.6]{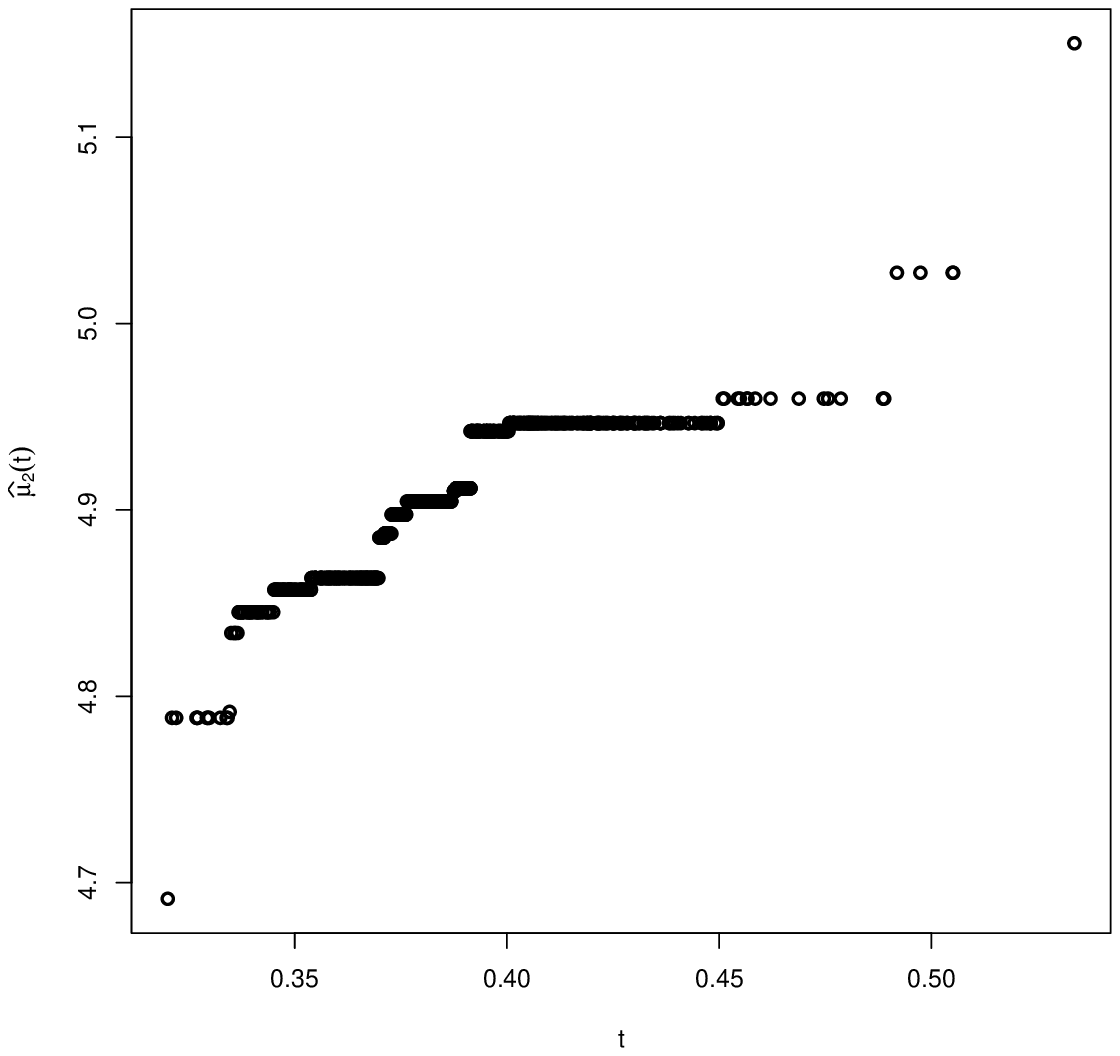}}
  \caption{Plot of $\widehat \mu_2(t)$ versus $t$ based on the UKBiobank data.}\label{estmu}
\end{figure}

In Figure \ref{estmu}, the estimated regression function $\widehat\mu_2(t)=\widehat\mu_{2,\widehat\beta}(t)$ in \eqref{eq-method-6},
 derived from the UK Biobank data, is presented. What is evident from this graph is that the relationship between the variables is not linear. This observation suggests that there is a notable difference or shift between the data from the UK Biobank and the ARIC study. Attempting to simply merge or combine these datasets by assuming they are drawn from identical distributions may lead to biased or inaccurate results.

The non-linearity in the estimated regression function suggests that there are likely different patterns or relationships between the variables in the two datasets. It's crucial for researchers to account for these differences and potential biases when conducting analyses or making comparisons between the two datasets. This recognition of dissimilarity can guide the use of more appropriate statistical methods and modeling approaches to address the unique characteristics of each dataset.


\section{Discussion} \label{Section-Dis}

In machine learning literature, distribution shifts between the training and testing data are a very common problem. Common distribution shifts include covariate shift and prior probability shift. Dealing with covariate shift is a relatively easy problem since the conditional distribution of $Y$ given $X$ remains the same in both datasets. However, it poses a real challenging problem in the presence of prior probability shift since the conditional density changes from $f(y|{\bm x})$ to $\pi(y)f(y|{\bm x})/\int\pi(y)f(y|{\bm x})dy$, where $\pi(y)$ is the shift function or selection bias function. This predictive density generally depends on the shift function $\pi(y)$. Without correcting this selection bias, we end up with a detrimented model.

In this paper, we have introduced a shape-restricted transfer learning approach that effectively merges training and testing data to enhance parameter estimation. This method integrates prior knowledge concerning shape restrictions seamlessly with various practical applications. The challenge of learning monotonic functions has been acknowledged for a long time, particularly when striving for top-tier performance on real-world issues that demand greater flexibility and the assurance of monotonicity across multiple features.

While the monotone single index model has been extensively discussed in statistical and econometric literature over the years, significant breakthroughs in semiparametric maximum likelihood estimation of index coefficients were only recently achieved, particularly by \citet{Groeneboom2018}. Notably, their work demonstrated that the monotone shape-restricted maximum likelihood of index coefficients is nearly root-$n$ consistent. Moreover, they cleverly constructed a root-$n$ zero-cross estimator using the estimated shape function, adding to the potential of this area of research.

The contributions of this paper open up new avenues for future research in shape-restricted transfer learning and monotonic function estimation, paving the way for improved solutions to various real-world challenges. By building upon these foundations, researchers can further enhance the understanding of complex datasets and unlock even greater potential for practical applications. The findings presented here represent a valuable step forward in the field of statistical and econometric modeling, and we hope they inspire further investigation and innovation in shape-constrained parameter estimation. As more researchers delve into this topic, we anticipate a deeper understanding of the theoretical and practical aspects, ultimately leading to more robust and accurate solutions in various domains.

There are multiple related topics that can be studied subsequently. Firstly, we assumed that the conditional density of $Y$ given $X = \bm{x}$ between the training and testing data satisfied the relationship $f^*(y|\bm{x}) \propto \pi(y) f(y|\bm{x})$, and that the effects of $X$ on $Y$ were linear through $X^T\beta$ with $\beta$ being a common parameter between the training and testing data. It might be of interest to consider other relationships between the conditional densities, such as those explored in \citet{caiwei2021} and \citet{Maity2021}, and explore the possibility that $\beta$ might differ across different datasets.
Secondly, our method in this paper is applicable when the dimension $p$ of $X$ is low. We expect that within the framework of this paper, we can extend our method to consider datasets with high-dimensional covariates \citep{Li2022, Tian2021}.
Thirdly, we may also extend our method to consider survival data with the response $Y$ being censored.
Lastly, in practice, the covariates of the training data and the testing data may consist of different variables. It could be of interest to consider the transfer learning method on such data.

\section*{Appendix: Sketch of the Proof for Part (3) of Theorems  \ref{theorem-lambda-positive} and \ref{theorem-lambda-0}}
\allowdisplaybreaks

For space limitations, we provide a blueprint of the proof for Part (3) of Theorems \ref{theorem-lambda-positive} and \ref{theorem-lambda-0}. The detailed proof for Parts (1) and (2) of these theorems is presented in the supplementary material.

Besides the notations in Sections \ref{Section-Method} and \ref{section-asymp}, we need the following notations. Denote by $F_{X_1, Y_1}(\bm{x}, y)$ and $\FF_{X_1,Y_1}(\bm{x}, y)$, respectively, the c.d.f. of $(X_1, Y_1)$ and the empirical c.d.f. of $\{(X_{1,i}, Y_{1,i}\}_{i=1}^n$, the observations of the training data. Similarly, we can define $F_{X_2, Y_2}(\bm{x}, y)$, $\FF_{X_2,Y_2}(\bm{x}, y)$, $F_{X_1}(\bm{x})$, and  $F_{X_2}(\bm{x})$. Let
\begin{eqnarray}
\psi(\beta) = \lambda \psi_1(\beta) + (1-\lambda) \psi_2(\beta), \label{eq-S-1-21}
\end{eqnarray}
with
\begin{eqnarray*}
\psi_1(\beta) &=& \int \frac{\mu_1'(\bm{x}^T\beta)}{V(\bm{x}^T \beta)} \bm{x}\left\{\mu_1(\bm{x}^T\beta_0) - \mu_1 (\bm{x}^T\beta) \right\} d F_{X_1}(\bm{x}), \\
\psi_2(\beta) &=&  \int \bm{x} \left\{\mu_{2,0}(\bm{x}^T\beta_0) - \mu_{2,\beta}(\bm{x}^T \beta) \right\} d F_{X_2}(\bm{x}). \label{eq-S-1-22}
\end{eqnarray*}
They are  the population versions of $\psi_{n,m}(\cdot)$, $\psi_{1,n}(\cdot)$, and $\psi_{2, m}(\cdot)$, respectively defined by \eqref{eq-method-8}, \eqref{eq-method-4}, and \eqref{eq-method-7}. Recalling the definition of $\varphi_\beta(t) = E(X_2|X_2^T\beta = t)$, and for any piece-wise constant function $\mu(\cdot)$ with finitely many jumps $\tau_1 <\tau_2 < \ldots$, we define
\begin{eqnarray}
\bar{\varphi}_{\beta, \mu}(t) = \left\{\begin{array}{ll} \varphi_\beta(\tau_i), & \text{if } \mu_{2,\beta}(t)> \mu(\tau_i), \text{ for all } t \in [\tau_i, \tau_{i+1}) \\ \varphi_\beta(s), & \text{if } \mu_{2,\beta}(s) = \mu(\tau_i), \text{ for some } s\in [\tau_i, \tau_{i+1}) \\ \varphi_\beta(\tau_{i+1}), & \text{if } \mu_{2,\beta}(t) < \mu(\tau_i), \text{ for all } t\in [\tau_i, \tau_{i+1})  \end{array} \right., \label{eq-S-1-6-added}
\end{eqnarray}
which is a constant function for each $i$ and $t \in [\tau_i, \tau_{i+1})$.

We first present two lemmas needed in our subsequent discussion. 

\begin{lemma} \label{lemma-initial}
Assume Conditions \ref{Condition-1}, \ref{Condition-5}, \ref{Condition-6}, \ref{Condition-7}, and \ref{Condition-8}. We have the following:
\begin{itemize}
    \item[(1)] $\psi(\beta_0) = 0$;
    \item[(2)] $\psi_1'(\beta_0) = \left.\frac{\partial \psi_1(\beta)}{\partial \beta}\right|_{\beta= \beta_0}$ exists with rank = $p$; $\psi_2'(\beta_0) = \left.\frac{\partial \psi_2(\beta)}{\partial \beta}\right|_{\beta= \beta_0}$ exists with rank = $p-1$;
    \item[(3)] We have
    \begin{eqnarray*}
    \psi_1(\beta) &=& \psi_1'(\beta_0)\cdot (\beta- \beta_0) + o(\beta- \beta_0)\\ 
    \psi_2(\beta) &=& \psi_2'(\beta_0)\cdot(\beta - \beta_0) + o(\beta - \beta_0). \label{eq-S-2-28}
    \end{eqnarray*}
\end{itemize}
\end{lemma}

\begin{lemma} \label{psi-B-rank} 
Assume Conditions \ref{Condition-5}, \ref{Condition-8}, and $\beta_{0,1}\neq 0$. For any $\lambda \in [0, 1)$, let 
\begin{eqnarray*}
B(\beta_0) =  \left( \begin{matrix} -\beta_{0,2:p}^T/\beta_{0,1} \\ I_{p-1} \end{matrix}\right), \label{eq-S-2-35}
\end{eqnarray*}
then $\psi_2'(\beta_0)_{-1} \cdot B(\beta_0)$ has rank $p-1$.  
\end{lemma}

Lemma \ref{lemma-initial} is Lemma 8, and Lemma \ref{psi-B-rank} is Lemma 10 in the supplementary material.

Consider an estimator $\check{\beta}$ of $\beta$ that satisfies 
\begin{eqnarray}
\check{\beta} - \beta_0 = o_p(1). \label{eq-S-4-5}
\end{eqnarray}
We derive the asymptotic properties of 
\begin{eqnarray}
\psi_{n,m}(\check{\beta}) = \lambda_{n,m} \psi_{1,n}(\check{\beta}) + (1-\lambda_{n,m}) \psi_{2,m}(\check{\beta}), \label{eq-S-4-6}
\end{eqnarray}
where $\psi_{1,n}(\cdot)$ and $\psi_{2,m}(\cdot)$ are defined by \eqref{eq-method-4} and \eqref{eq-method-7}, respectively. 
Next, we consider the asymptotic properties of $\psi_{1,n}(\check{\beta})$ and  $\psi_{2,m}(\check{\beta})$ separately. For $\psi_{2,m}(\check{\beta})$, we have
\begin{eqnarray*}
\psi_{2,m}(\check\beta) &=& \frac{1}{m}\sum_{j=1}^m X_{2,j}\left\{Y_{2,j}-\widehat \mu_{2,\check \beta}\left(X_{2,j}^T\check \beta\right)\right\}\\
& = & \int \bm{x} \left\{y - \widehat \mu_{2,\check \beta}\left(\bm{x}^T\check \beta\right)\right\} d \FF_{X_2,Y_2}(\bm{x}, y) \\
&=& \int \left\{\bm{x} - \bar \varphi_{\check \beta, \widehat \mu_{2, \check \beta}}\left(\bm{x}^T \check \beta \right) \right\} \left\{y - \widehat \mu_{2,\check \beta}\left(\bm{x}^T\check \beta\right)\right\} d \FF_{X_2,Y_2}(\bm{x}, y)\\
&=& J_1 + J_2, \label{eq-S-4-7}
\end{eqnarray*}
where 
\begin{eqnarray}
J_1 &=& \int \left\{\bm{x} - \varphi_{\check \beta}\left(\bm{x}^T \check \beta \right) \right\} \left\{y - \widehat \mu_{2,\check \beta}\left(\bm{x}^T\check \beta\right)\right\} d \FF_{X_2,Y_2}(\bm{x}, y) \label{eq-S-4-8-added},\\
J_2 &=& \int \left\{\varphi_{\check \beta}\left(\bm{x}^T \check \beta \right) -  \bar \varphi_{\check \beta, \widehat \mu_{2, \check \beta}}\left(\bm{x}^T \check \beta \right)\right\} \left\{y - \widehat \mu_{2,\check \beta}\left(\bm{x}^T\check \beta\right)\right\} d \FF_{X_2,Y_2}(\bm{x}, y). \label{eq-S-4-8}
\end{eqnarray}
In the derivation above, we have used the fact that 
\begin{eqnarray*}
\int \bar \varphi_{\check \beta, \widehat \mu_{2, \check \beta}}\left(\bm{x}^T \check \beta \right) \left\{y - \widehat \mu_{2,\check \beta}\left(\bm{x}^T\check \beta\right)\right\} d \FF_{X_2,Y_2}(\bm{x}, y) = 0. \label{eq-S-4-9}
\end{eqnarray*}
This is because $\bar \varphi_{\check \beta, \widehat \mu_{2, \check \beta}}\left(\bm{x}^T \check \beta \right)$ is a piece-wise constant function with the same jumps as $\widehat \mu_{2,\check \beta}(t)$. Recall the definition of $\widehat \mu_{2,\check \beta}(t)$: it is the slope of the greatest convex minorant of the corresponding cusum diagram, based on the values of $Y_{2,j}$ in the order of $X_{2,j}^T\check \beta$ for $j=1,\ldots, m$. Therefore, on every constant piece $[\tau_i, \tau_{i+1})$, 
\begin{eqnarray*}
\int_{\bm{x}^T \check \beta \in [\tau_i, \tau_{i+1})}  \left\{y - \widehat \mu_{2,\check \beta}\left(\bm{x}^T\check \beta\right)\right\} d \FF_{X_2,Y_2}(\bm{x}, y) = 0. \label{eq-S-4-10}
\end{eqnarray*}

We summarize the asymptotic properties of $J_1$ and $J_2$ in two separate lemmas: Lemmas \ref{lemma-asym-J-1} and \ref{lemma-asym-J-2} below.  
Lemma \ref{lemma-asym-J-1} establishes the asymptotic properties of $J_1$, while Lemma \ref{lemma-asym-J-2} considers those of $J_2$. They are Lemmas 12 and 13 in the supplementary material, and  their proofs are also provided there. 
Combining these two lemmas, we immediately obtain 
\begin{equation}
\psi_{2,m}(\check \beta) = \psi_2(\check \beta) + \int \left\{\bm{x} - \varphi_{\beta_0}\left(\bm{x}^T \beta_0 \right) \right\} \left\{y - \mu_{2,0}\left(\bm{x}^T\beta_0\right)\right\}d\FF_{X_2,Y_2}(\bm{x}, y) + o_p\left(m^{-1/2}\right). \label{eq-S-4-10-added}
\end{equation}

\begin{lemma} \label{lemma-asym-J-1}
Assume Conditions \ref{Condition-1}--\ref{Condition-8}, and \eqref{eq-S-4-5}. Referring to $J_1$ as defined by \eqref{eq-S-4-8-added}, we have 
\begin{equation}
J_1 = \psi_2(\check \beta) + \int \left\{\bm{x} - \varphi_{\beta_0}\left(\bm{x}^T \beta_0 \right) \right\} \left\{y - \mu_0\left(\bm{x}^T\beta_0\right)\right\}d\FF_{X_2,Y_2}(\bm{x}, y) + o_p\left(m^{-1/2}\right). \label{eq-S-4-11-added}
\end{equation}
\end{lemma}

\begin{lemma}\label {lemma-asym-J-2}
Assume Conditions \ref{Condition-1}--\ref{Condition-8}, and \eqref{eq-S-4-5}.  Referring to $J_2$ as defined in \eqref{eq-S-4-8}, we have
\begin{eqnarray*}
J_2 = o_p\left(m^{-1/2}\right). 
\end{eqnarray*}
\end{lemma}

With a similar but simpler development as \eqref{eq-S-4-10-added} (details can be found in the supplementary material), we obtain 
\begin{eqnarray}
\psi_{1,n} = \psi_1(\check \beta) + \int \frac{\mu_1'(\bm{x}^T\beta_0)}{V(\bm{x}^T\beta_0)} \bm{x} \left\{ y - \mu_1(\bm{x}^T \beta_0) \right\} d \FF_{X_1, Y_1}(\bm{x}, y) + o_p\left(n^{-1/2}\right). \label{eq-S-4-60}
\end{eqnarray}
Combining \eqref{eq-S-4-6}, \eqref{eq-S-4-10-added}, and \eqref{eq-S-4-60}, we have
\begin{eqnarray}
\psi_{n,m}(\check \beta) &=& \lambda_{n,m}\psi_1(\check \beta) + (1-\lambda_{n,m}) \psi_2(\check\beta) \nonumber \\
&& + \lambda_{n,m} \int \frac{\mu_1'(\bm{x}^T\beta_0)}{V(\bm{x}^T\beta_0)} \bm{x} \left\{ y - \mu_1(\bm{x}^T \beta_0) \right\} d \FF_{X_1, Y_1}(\bm{x}, y) \nonumber \\
&&+ (1-\lambda_{n,m}) \int \left\{\bm{x} - \varphi_{\beta_0}\left(\bm{x}^T \beta_0 \right) \right\} \left\{y - \mu_{2,0}\left(\bm{x}^T\beta_0\right)\right\}d\FF_{X_2,Y_2}(\bm{x}, y) \nonumber \\
&&+ \lambda_{n,m} \cdot  o_p\left( n^{-1/2}\right) + (1-\lambda_{n,m}) \cdot  o_p\left(m^{-1/2}\right) \nonumber \\
&=& \lambda_{n,m}\psi_1(\check \beta) + (1-\lambda_{n,m}) \psi_2(\check\beta) \nonumber \\
&& + \lambda_{n,m} \int \frac{\mu_1'(\bm{x}^T\beta_0)}{V(\bm{x}^T\beta_0)} \bm{x} \left\{ y - \mu_1(\bm{x}^T \beta_0) \right\} d \FF_{X_1, Y_1}(\bm{x}, y) \nonumber \\
&&+ (1-\lambda_{n,m}) \int \left\{\bm{x} - \varphi_{\beta_0}\left(\bm{x}^T \beta_0 \right) \right\} \left\{y - \mu_{2,0}\left(\bm{x}^T\beta_0\right)\right\}d\FF_{X_2,Y_2}(\bm{x}, y) \nonumber \\
&&+  o_p\left((n+m)^{-1/2}\right).\label{eq-S-4-61}
\end{eqnarray}

With the same arguments as those in Section 4 of the supplementary material, we have consistent estimators of $\widehat{\beta}$: $\widehat{\beta}^L$ and $\widehat{\beta}^U$, and $0 \leq \alpha_{n,m} \leq 1$ (in the proof for Part (1) of Theorems \ref{theorem-lambda-positive} and \ref{theorem-lambda-0}). Furthermore, for $\lambda > 0$, based on \eqref{eq-S-4-61} and Lemma \ref{lemma-initial}, we have
\begin{eqnarray}
0&=&\alpha_{n,m} \psi_{n,m}\left(\widehat \beta^L\right) + (I_p-\alpha_{n,m}) \psi_{n,m}\left(\widehat \beta^U\right) \nonumber \\
& = & \alpha_{n,m} \left\{\lambda_{n,m}\psi_1\left(\widehat \beta^L\right) + (1-\lambda_{n,m}) \psi_2\left(\widehat \beta^L\right)\right\} \nonumber \\ 
&& + (I_p-\alpha_{n,m}) \left\{\lambda_{n,m}\psi_1\left(\widehat \beta^U\right) + (1-\lambda_{n,m}) \psi_2\left(\widehat \beta^U\right)\right\} \nonumber \\
&& + \lambda_{n,m} \int \frac{\mu_1'(\bm{x}^T\beta_0)}{V(\bm{x}^T\beta_0)} \bm{x} \left\{ y - \mu_1(\bm{x}^T \beta_0) \right\} d \FF_{X_1, Y_1}(\bm{x}, y) \nonumber \\
&&+ (1-\lambda_{n,m}) \int \left\{\bm{x} - \varphi_{\beta_0}\left(\bm{x}^T \beta_0 \right) \right\} \left\{y - \mu_{2,0}\left(\bm{x}^T\beta_0\right)\right\}d\FF_{X_2,Y_2}(\bm{x}, y) \nonumber \\
&&+  o_p\left((n+m)^{-1/2}\right) \nonumber \\
&=& A(\beta_0) \left(\widehat \beta - \beta_0\right)  + \lambda_{n,m} \int \frac{\mu_1'(\bm{x}^T\beta_0)}{V(\bm{x}^T\beta_0)} \bm{x} \left\{ y - \mu_1(\bm{x}^T \beta_0) \right\} d \FF_{X_1, Y_1}(\bm{x}, y) \nonumber \\
&&+ (1-\lambda_{n,m}) \int \left\{\bm{x} - \varphi_{\beta_0}\left(\bm{x}^T \beta_0 \right) \right\} \left\{y - \mu_{2,0}\left(\bm{x}^T\beta_0\right)\right\}d\FF_{X_2,Y_2}(\bm{x}, y) \nonumber \\
&&+  o_p\left((n+m)^{-1/2}\right) + o_p\left(\widehat \beta - \beta_0\right) \nonumber \\
&=& A(\beta_0) \left(\widehat \beta - \beta_0\right) + \lambda_{n,m}^{1/2} (n+m)^{-1/2} G_{1,n} + (1-\lambda_{n,m})^{1/2} (n+m)^{-1/2} G_{2,m} \nonumber \\
&& +  o_p\left((n+m)^{-1/2}\right) + o_p\left(\widehat \beta - \beta_0\right), \label{eq-S-4-62}
\end{eqnarray}
where $A(\beta_0) = \lambda \psi_1'(\beta_0) + (1-\lambda) \psi_2'(\beta_0)$,
\begin{eqnarray*}
G_{1,n} &=& \sqrt{n} \int \frac{\mu_1'(\bm{x}^T\beta_0)}{V(\bm{x}^T\beta_0)} \bm{x} \left\{ y - \mu_1(\bm{x}^T \beta_0) \right\} d \FF_{X_1, Y_1}(\bm{x}, y),\\
G_{2,m} &=& \sqrt{m}\int \left\{\bm{x} - \varphi_{\beta_0}\left(\bm{x}^T \beta_0 \right) \right\} \left\{y - \mu_{2,0}\left(\bm{x}^T\beta_0\right)\right\}d\FF_{X_2,Y_2}(\bm{x}, y). \label{eq-S-4-63}
\end{eqnarray*}
Based on Conditions \ref{Condition-3}--\ref{Condition-7} and the Central Limit Theorem, we have
\begin{eqnarray*}
&G_{1,n} \to& N\left(0, \Sigma_1 \right), \\
\text{and} & G_{2,m} \to& N\left(0, \Sigma_2\right), \label{eq-S-4-64}
\end{eqnarray*}
in distribution, where 
\begin{eqnarray*}
\Sigma_1 &=& \text{var}\left[ \frac{\mu_1'(X_1^T\beta_0)}{V(X_1^T\beta_0)} X_1 \left\{ Y_1 - \mu_1(X_1^T \beta_0) \right\} \right]\\
\Sigma_2 &=& \text{var}\left[ \left\{X_2 - \varphi_{\beta_0}\left(X_2^T \beta_0 \right) \right\} \left\{Y_2 - \mu_{2,0}\left(X_2^T\beta_0\right)\right\} \right]. 
\end{eqnarray*}
We, therefore, conclude that the c.d.f. of $(G_{1,n}^T, G_{2,m}^T)^T$ is given by:
\begin{eqnarray*}
&&P\left( (G_{1,n}^T, G_{2,m}^T)^T \leq (g_1^T, g_2^T)^T \right) \\ 
& = & P\left(G_{1,n} \leq g_1\right) P\left( G_{2,m} \leq g_2 \right) \\
&\to & \Phi_{\Sigma_1}(g_1) \Phi_{\Sigma_2}(g_2), \label{eq-S-4-65}
\end{eqnarray*}
for any $g_1, g_2 \in \RR^p$, where $\Phi_{\Sigma}(\cdot)$ denotes the c.d.f. for the $N(0, \Sigma)$ distribution. This implies
\begin{eqnarray}
\left(\begin{matrix} G_{1,n} \\ G_{2,m} \end{matrix} \right) \to N\left(0, \left(\begin{matrix} \Sigma_1 & 0 \\ 0 & \Sigma_2 \end{matrix} \right) \right), \label{eq-S-4-66}
\end{eqnarray}
in distribution. Combining \eqref{eq-S-4-62} with \eqref{eq-S-4-66}, applying Delta method, and based on the fact that $\lambda_{n,m} \to \lambda$, we conclude 
\begin{eqnarray*}
\sqrt{n+m} \left( \widehat \beta - \beta_0 \right) \to N\left(0, A_0(\beta_0)^{-1}\left\{ \lambda \Sigma_1 + (1-\lambda) \Sigma_2 \right\} A_0^{-1}(\beta_0) \right), \label{eq-S-4-67}
\end{eqnarray*}
in distribution, where 
\begin{eqnarray*}
A(\beta_0)  &=& -\lambda \psi_1'(\beta_0) - (1-\lambda) \psi_2'(\beta_0) \\
&=&  \lambda E\left[ \frac{\left\{\mu_1'(X_1^T\beta_0)\right\}^2}{V(X_1^T \beta_0)} X_1 X_1^T \right] \\
&& + (1-\lambda)E\left[ \left\{X_2- \varphi_{\beta_0}(X_2^T \beta_0)\right\}\left\{X_2 - \varphi_{\beta_0}(X_2^T \beta_0)\right\}^T \mu_{2,0}'\left(X_2^T\beta_0 \right)   \right]. \label{eq-S-4-68}
\end{eqnarray*}

For $\lambda = 0$, since $\widehat \beta$ is a zero crossing of $\psi_{n,m}(\beta)_{2:p}$,  based on \eqref{eq-S-4-61} and Lemma \ref{lemma-initial}, we have 
\begin{eqnarray}
0&=&\alpha_{n,m} \psi_{n,m}\left(\widehat \beta^L\right)_{2:p} + (I_{p-1}-\alpha_{n,m}) \psi_{n,m}\left(\widehat \beta^U\right)_{2:p} \nonumber \\
& = & \alpha_{n,m} \left\{\lambda_{n,m}\psi_1\left(\widehat \beta^L\right) + (1-\lambda_{n,m}) \psi_2\left(\widehat \beta^L\right)\right\}_{2:p} \nonumber \\ 
&& + (I_p-\alpha_{n,m}) \left\{\lambda_{n,m}\psi_1\left(\widehat \beta^U\right) + (1-\lambda_{n,m}) \psi_2\left(\widehat \beta^U\right)\right\}_{2:p} \nonumber \\
&& + \lambda_{n,m} \int \frac{\mu_1'(\bm{x}^T\beta_0)}{V(\bm{x}^T\beta_0)} \bm{x}_{2:p} \left\{ y - \mu_1(\bm{x}^T \beta_0) \right\} d \FF_{X_1, Y_1}(\bm{x}, y) \nonumber \\
&&+ (1-\lambda_{n,m}) \int \left\{\bm{x}_{2:p} - \varphi_{\beta_0}\left(\bm{x}^T \beta_0 \right)_{2:p} \right\} \left\{y - \mu_{2,0}\left(\bm{x}^T\beta_0\right)\right\}d\FF_{X_2,Y_2}(\bm{x}, y) \nonumber \\
&&+  o_p\left((n+m)^{-1/2}\right) \nonumber \\
&=& \lambda_{n,m} \psi_1(\widehat \beta)_{2:p} + (1-\lambda_{n,m}) \psi_2(\widehat \beta)_{2:p} \nonumber \\
&& + \lambda_{n,m}^{1/2} (n+m)^{-1/2} \widetilde G_{1,n} + (1-\lambda_{n,m})^{1/2} (n+m)^{-1/2} \widetilde G_{2,m} \nonumber \\
&&+  o_p\left((n+m)^{-1/2}\right) \nonumber \\
&=& - A(\beta_0)_{2:p} B(\beta_0) \left(\widehat \beta_{2:p} - \beta_{0, 2:p}\right)  \nonumber \\
&& + \lambda_{n,m}^{1/2} (n+m)^{-1/2} \widetilde G_{1,n} + (1-\lambda_{n,m})^{1/2} (n+m)^{-1/2} \widetilde G_{2,m} \nonumber \\
&&+  o_p\left((n+m)^{-1/2}\right) + o_p\left(\widehat \beta_{2:p} - \beta_{0, 2:p}\right)\nonumber \\
&=& \psi_2'(\beta_0)_{2:p} B(\beta_0) \left(\widehat \beta_{2:p} - \beta_{0, 2:p}\right)  \nonumber \\
&& + \lambda_{n,m}^{1/2} (n+m)^{-1/2} \widetilde G_{1,n} + (1-\lambda_{n,m})^{1/2} (n+m)^{-1/2} \widetilde G_{2,m} \nonumber \\
&&+  o_p\left((n+m)^{-1/2}\right) + o_p\left(\widehat \beta_{2:p} - \beta_{0, 2:p}\right), \label{eq-S-4-69}
\end{eqnarray}
where 
\begin{eqnarray*}
\widetilde G_{1,n} &=& \sqrt{n} \int \frac{\mu_1'(\bm{x}^T\beta_0)}{V(\bm{x}^T\beta_0)} \bm{x}_{2:p} \left\{ y - \mu_1(\bm{x}^T \beta_0) \right\} d \FF_{X_1, Y_1}(\bm{x}, y), \\
\widetilde G_{2,m} &=& \sqrt{m}\int \left\{\bm{x}_{2:p} - \varphi_{\beta_0}\left(\bm{x}^T \beta_0 \right)_{2:p} \right\} \left\{y - \mu_{2,0}\left(\bm{x}^T\beta_0\right)\right\}d\FF_{X_2,Y_2}(\bm{x}, y),\\
 B(\beta_0) &=&  \left( \begin{matrix} -\beta_{0,2:p}^T/\beta_{0,1} \\ I_{p-1} \end{matrix}\right).
\label{eq-S-4-70}
\end{eqnarray*}
Based on Lemma \ref{psi-B-rank}, $\psi_2'(\beta_0)_{2:p} B(\beta_0)$ is of full rank. Based on Conditions \ref{Condition-3}--\ref{Condition-7}, we have $\widetilde G_{1,n} = O_p(1)$, and
\begin{eqnarray}
\lambda_{n,m}^{1/2} (n+m)^{-1/2} \widetilde G_{1,n} &=& o_p\left((n+m)^{-1/2}\right) \nonumber \\
\widetilde G_{2,m} &\to&  N(0, \widetilde \Sigma_2), \label{eq-S-4-71}
\end{eqnarray}
with
\begin{eqnarray*}
\widetilde \Sigma_2 &=& \text{var}\left[ \left\{(X_2)_{2:p} - \varphi_{\beta_0}\left(X_2^T \beta_0 \right)_{2:p} \right\} \left\{Y_2 - \mu_{2,0}\left(X_2^T\beta_0\right)\right\} \right] = \left(\Sigma_2 \right)_{-1,-1}. \label{eq-S-4-72}
\end{eqnarray*}

Combining \eqref{eq-S-4-69} with \eqref{eq-S-4-71} leads to 
\begin{eqnarray*}
\sqrt{n+m} \left(\widehat \beta_{2:p} - \beta_{0, 2:p} \right)\to \left\{\widetilde A(\beta_0) \right\}^{-1} N(0, \widetilde \Sigma_2),
\end{eqnarray*}
in distribution, where with straightforward evaluations, we have,
\begin{eqnarray*}
\widetilde A(\beta_0)  &=& -\psi_2'(\beta_0)_{2:p} B(\beta_0) \\
&=& E\left[ \left\{X_2 - \varphi_{\beta_0}(X_2^T \beta_0)\right\}_{2:p}\left\{X_2 - \varphi_{\beta_0}(X_2^T \beta_0)\right\}_{2:p}^T \mu_{2, 0}'\left(X_2^T\beta_0 \right)   \right] \nonumber \\ 
&&\times \left(\beta_{0,2:p} \beta_{0,2:p}^T + I_{p-1}\right). 
\end{eqnarray*}
We complete the proof for Part (3) of Theorems \ref{theorem-lambda-positive}
and \ref{theorem-lambda-0}.

\bibliographystyle{apalike}
\bibliography{Reference}

\end{document}